# Monte Carlo uncertainty estimation for wavelength calibration

**Thang H.L., Nguyen D.D., Dung.D.N.*

*Vietnam metrology institute*

*8 Hoang Quoc Viet street, Cau Giay district, Hanoi city, Vietnam*

** thanglh@vmi.gov.vn*

**Abstract**: CIPM published the Supplement I for GUM in 2008 as not only an alternative approach to estimate the uncertainty for a given calibration measurement but also as a proper uncertainty estimation one, whenever any of the conditions imposed in GUM which must be satisfied does not hold [1, 2]. Before the introduction of the new approach in the Supplement I, namely Monte Carlo (MC) method, the GUM rules have been always applied even if in cases where the mentioned conditions were not fulfilled. After or even before the official introduction of this MC method, a number of published papers in uncertainty estimation by using MC method had been shown up, giving more insight for the ways the uncertainties estimated and also for the specific calibration measurements under investigation themselves [3 - 16]. However, in most of those published papers, the application conditions required before a method selection should be analyzed, in fact, has been ignored. In our uncertainty study for wavelength calibration by using beat frequency technique, the application conditions have not been ignored but analyzed carefully and it was found that a GUM application condition was unsatisfactory. However, all the MC method conditions for this calibration problem have been well fulfilled. A computer program was then coded [17] following the steps in the Supplement I. The resultant outputs showed difference in statistical nature and suggested to replace the MC method to the GUM one as soon as possible for this case.

## I Introduction:

CIPM published the Supplement I for GUM in 2008 as not only an alternative approach to estimate the uncertainty for a given calibration measurement but also as a proper uncetainty estimation approach, whenever any of the conditions required must be satisfied in order to be able to follow the steps in GUM, does not hold [1, 2]. Before the introduction of the new approach in the Supplement I, namely Monte Carlo (MC) method, the GUM rules have been always applied even if in cases the mentioned conditions were not fulfilled (perhaps, because there was no other choice for people to calculate this parameter then). After or even before the official introduction of this MC method, a number of published papers in uncertainty estimation by using MC method have been shown up, giving more insight for the way the uncertainties estimated and also for the specific calibration measurements under investigation themselves [3 - 16]. In particular, in the case of wavelength calibration using beat frequency technique, it is apparent that some of the required conditions is not fulfilled for GUM to be used as seen in the following examination. However, all the conditions required in the MC method for this calibration problem have been well fulfilled. Therefore the impossibility of using the GUM method for uncertainty estimation becomes clear, and the MC application is now becoming the only right option.

To numerically illustrate for the argument mentioned above, we carried out an uncertainty estimation by using the MC method for a wavelength calibration measurement by using beat frequency technique for an HP laser source which has been long time served as a wavelength standard in the laboratory of length measurement at Vietnam Metrology Institute (VMI). The resulted uncertainties estimated from those two approaches have been seen quite different in statistical natures as will be seen in the following sections.

## II. Calibration measurement setup and modeling

The wavelength calibration setup is of a typical beat frequency technique. The standard laser and the laser under test (LUT) are fixed on an anti vibration optical table with optical mirrors, splitter, electric optical modulation etc. as shown in the diagram in figure 1 [18]. The beat signal is led to the frequency counter. The spectrum analyzer is used for rough tuning the beat. The beat signal will be then transferred to a computer and the data is processed here with the assistance of the Laser call 3.0 software [19].

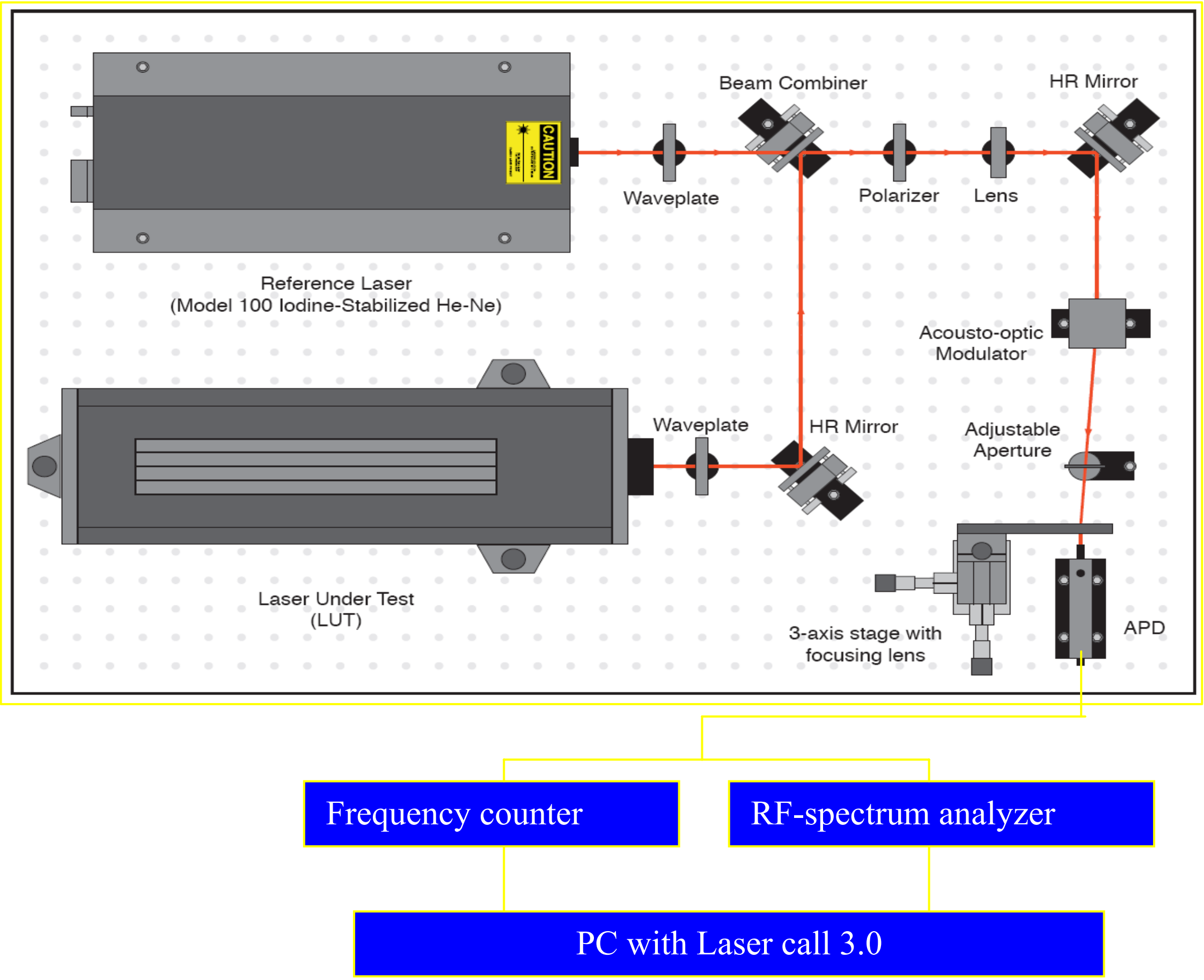

*Figure1. Beat frequency technique setup used for wavelength calibration.*

The standard laser is an iodine frequency stabilized He-Ne one, model Winter 100. This laser source generates a 633 nm laser beam, with the relative uncertainty $2.5 \times 10^{-11}$. The 633 nm LUT is a laser source made by HP [19, 20].

In the wavelength calibration measurement, the often used mathematical model has the form:

$$Y = F(X) = X(1) + X(2) + X(3) \quad (1)$$

Where X are the influencing factors constitute of X(1) – standard frequency, X(2) – beat frequency and X(3) – division of the frequency counter. The measurand Y is the frequency of the LUT.

It is seen from the model equation (1) that this model equation is linear in its nature, and its first order partial derivatives with respect to the input independent variables are in existence. It is also straightforward to check the continuity of the equation (1) within the vicinity of the estimated input variables. The distribution of Y is also increasing and the density function is single-peaked etc. We can summarize the conditions which need to be satisfied when one would like to apply either approach in table 1. The satisfied condition will be marked as (X) otherwise there will be left blank.

*Table1. Application conditions need to be satisfied for a method to be applied in this calibration measurement.*

| GUM's conditions | GUM satisfaction | Supplement I - Monte Carlo method's conditions | MC satisfaction |
|---|---|---|---|
| the Welch - Satterthwaite formula is adequate for calculating the effective degrees of freedom associated with u(y) [GUM:1995 G.4.1], when one or more of the u(xi) has an associated degrees of freedom that is finite; | (X) | F(X) is continuous with respect to the elements X(i) of X in the neighbourhood of the best estimates x(i) of the X(i) | (X) |
| the Xi are independent when the degrees of freedom associated with the u(xi) are finite; | (X) | the distribution function for Y is continuous and strictly increasing | (X) |
| the PDF for Y can adequately be approximated by a Gaussian distribution or a scaled and shifted t-distribution. | | the PDF for Y is<br>1) continuous over the interval for which this PDF is strictly positive,<br>2) unimodal (single-peaked), | (X)<br>(X) |
| | | and<br>3) strictly increasing (or zero) to the left of the mode and strictly decreasing (or zero) to the right of the mode;<br>4) E(Y ) and V (Y ) exist;<br>5) a sufficiently large value of M is used. | <br>(X)<br><br><br><br>(X)<br>(X) |

After an examination of the required conditions as shown in table 1, it is apparent that the GUM method should not be used due to the unsatisfactory last condition. The reason is the dominant influencing factor of beat frequency has the probability density function of rectangular, leading to the non-student distribution of the output measurand values. As it was mentioned above, the GUM rules have been still used in the community, however, by the understanding of the authors, due to there has been no other approach around then. It seems very serious if then the estimated uncertainty of the wavelength would be applied to a next measurement with the assumption of its Gaussian/Student distribution as often misused unavoidable among the community. Fortunately, all the conditions required in the MC use for this problem have been seen satisfactory. So in this case, in order to properly estimate the uncertainty for the wavelength problem using the model (1), MC method becomes the only right option.

## III. Result and discussion

The values of those above stated influencing factors and their corresponding standard uncertainties and some other parameters of the laser sources stated above are given in table 2 as follows:

*Table2. Information of influencing factors: symbol, value, type of distribution, upper and lower limit, divisor, standard uncertainty and sensitivity. Dimension in GHz.*

| № | Influencing factors | Symbol | Value | Distribution | Lower limit | Upper limit | Divisor | $u_i$ | $c_i$ |
|---|---|---|---|---|---|---|---|---|---|
| 01 | Standard frequency | X(1) | 473612.3536 | Gaussian | - | - | 1 | 1.2E-05 | 1 |
| 02 | Beat frequency | X(2) | 0.16305987 | Rectangular | -0.0012623 | 0.00126226 | 1.73205 | 0.00073 | 1 |
| 03 | Division of counter | X(3) | 0 | Rectangular | -5E-12 | 5E-12 | 1.73205 | 2.9E-12 | 1 |

We have coded a computer program [17] following the steps given in the Supplement I. Take the inputs in table 2 to run the code we obtained the results represented in figure 3. The distributions of the three influencing factors used as the inputs for the Mote Carlo method have the forms in figure 2.

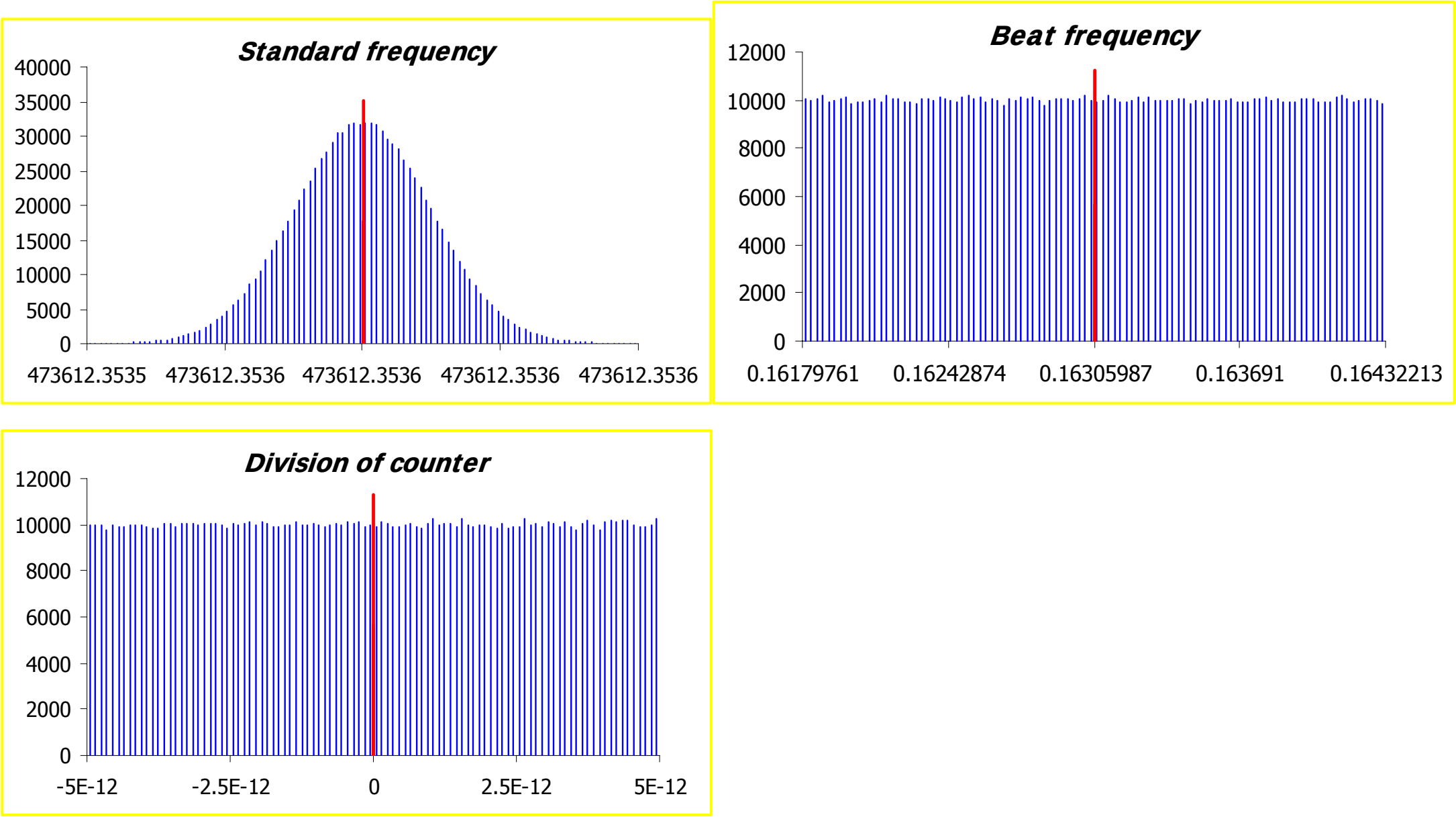

*Figure 2. Monte Carlo method to estimate the uncertainty: histograms of the input influencing factors. Horizontal axis is frequency in GHz, vertical is the appearance counting.*

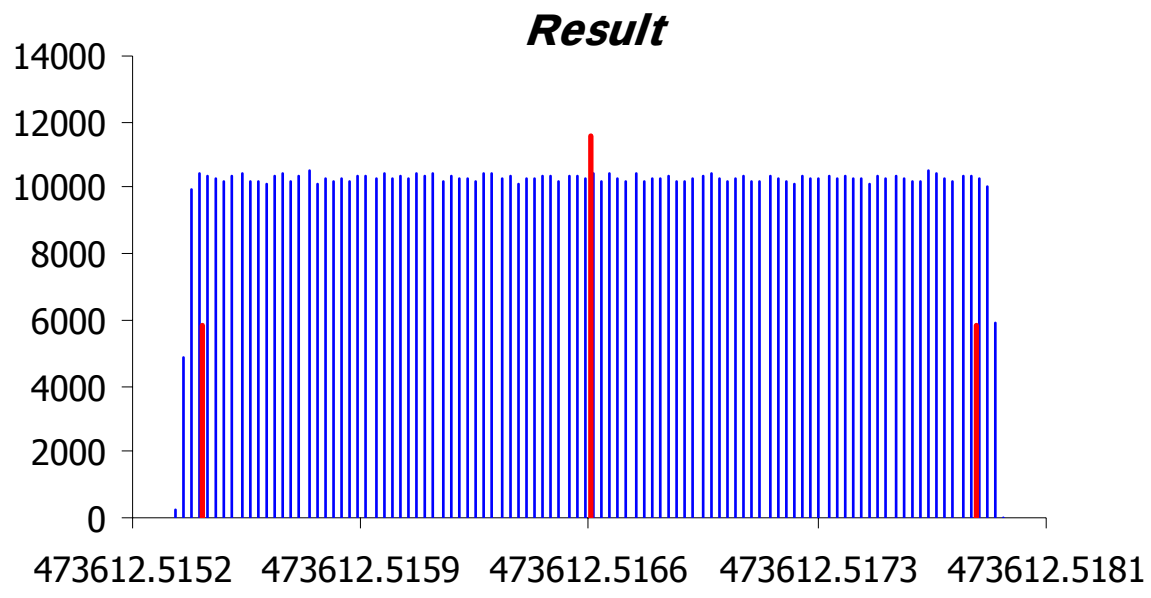

*Figure3. Monte Carlo method to estimate the uncertainty: the resultant output histogram. Horizontal axis is frequency in GHz, vertical is the appearance counting.*

From the resultant distribution for the measurand, the lower and upper limit of the range corresponding to the probability 95% has been obtained. The uncertainty has been calculated by the method suggested in GUM for the comparison purpose also (although GUM should not be applicable inhere). The estimates reached by the both ways have been tabulated in table 3.

*Table3. Uncertainty (probability 95%) and the limits calculated via MC and GUM method. Values are in GHz.*

| | |
|---|---|
| GUM uncertainty | 0.001458 |
| MC lower | -0.001199 |
| MC upper | 0.001199 |

It is clearly seen that the resultant probability density function is an approximately rectangular one, not a student or normal. This result agrees with the one obtained from the uncertainty budget observation itself. Furthermore, there is a difference also in the final estimated uncertainty.

## IV. Conclusion

Since the birth of GUM in 1993, a united way for uncertainty expression over the world metrology community has been established. However the application of this approach has not been scientifically proper in every case, but still people keep using it because there was no other option around. After the official introduction of the MC method in the Supplement I in 2008 from CIPM, the application conditions for the both GUM and MC methods have been considered more thoroughly. Consequently those conditions need to be examined before a decision on which method should be used will be made. In the wavelength calibration uncertainty estimation problem, we have shown that there was an unsatisfactory condition existed in GUM application, but all of the conditions in the MC approach are fulfilled. The MC resultant distribution has an un-simple form far off the Gaussian or Student one as commonly expected or conventioned in GUM previously. This difference suggests that all the conditions required in both methods must be analyzed carefully before a method selection is decided. In particular, the MC method needs to be adopted as soon as possible in the wavelength calibration in place of the GUM one, because the GUM one has been shown un-applicable, but the MC one.